\documentclass[format=acmsmall, review=false, screen=true]{acmart}

\usepackage{booktabs} % For formal tables

\usepackage{multirow}
\usepackage{subfigure}
\usepackage{color}

\usepackage[ruled]{algorithm2e} % For algorithms

\SetAlFnt{\small}
\SetAlCapFnt{\small}
\SetAlCapNameFnt{\small}
\SetAlCapHSkip{0pt}
\IncMargin{-\parindent}

\acmJournal{TOIT}
\begin{document}
% Title portion. Note the short title for running heads
\title[A Dynamic Service-Migration Mechanism]{A Dynamic Service-Migration Mechanism in Edge Cognitive Computing}
\author{Min Chen}
%\orcid{1234-5678-9012-3456}
\affiliation{%
  \institution{Huazhong University of Science and Technology}
  \city{Wuhan}
  \postcode{430074}
  \country{China}}
\email{minchen@ieee.org}
\author{Wei Li}
\affiliation{%
  \institution{Huazhong University of Science and Technology}
  \city{Wuhan}
  \postcode{430074}
  \country{China}}
\email{weili\_epic@hust.edu.cn}
\author{Giancarlo Fortino}
\affiliation{%
 \institution{University of Calabria}
 \city{Rende}
 \postcode{87036}
 \country{Italy}}
\email{g.fortino@unical.it}
\author{Yixue Hao}
\affiliation{%
  \institution{Huazhong University of Science and Technology}
  \city{Wuhan}
  \postcode{430074}
  \country{China}}
\email{yixuehao@hust.edu.cn}
\author{Long Hu}
\affiliation{%
  \institution{Huazhong University of Science and Technology}
  \city{Wuhan}
  \postcode{430074}
  \country{China}}
\email{longhu.cs@gmail.com}
\author{Iztok Humar}
\affiliation{%
  \institution{University of Ljubljana}
  \country{Slovenia}
}
\email{iztok.humar@fe.uni-lj.si}

\begin{abstract}
Driven by the vision of edge computing and the success of rich cognitive services based on artificial intelligence, a new computing paradigm, edge cognitive computing (ECC), is a promising approach that applies cognitive computing at the edge of the network. ECC has the potential to provide the cognition of users and network environmental information, and further to provide elastic cognitive computing services to achieve a higher energy efficiency and a higher Quality of Experience (QoE) compared to edge computing. This paper firstly introduces our architecture of the ECC and then describes its design issues in detail. Moreover, we propose an ECC-based dynamic service migration mechanism to provide an insight into how cognitive computing is combined with edge computing. In order to evaluate the proposed mechanism, a practical platform for dynamic service migration is built up, where the services are migrated based on the behavioral cognition of a mobile user. The experimental results show that the proposed ECC architecture has ultra-low latency and a high user experience, while providing better service to the user, saving computing resources, and achieving a high energy efficiency.
\end{abstract}

%\begin{CCSXML}
%<ccs2012>
%  <concept_id>10010520.10010553.10010562</concept_id>
%  <concept_desc>Computer systems organization~Embedded systems</concept_desc>
%  <concept_significance>500</concept_significance>
%  <concept_id>10010520.10010575.10010755</concept_id>
%  <concept_desc>Computer systems organization~Redundancy</concept_desc>
%  <concept_significance>300</concept_significance>
%  <concept_id>10010520.10010553.10010554</concept_id>
%  <concept_desc>Computer systems organization~Robotics</concept_desc>
%  <concept_significance>100</concept_significance>
%  <concept_id>10003033.10003083.10003095</concept_id>
%  <concept_desc>Networks~Network reliability</concept_desc>
%  <concept_significance>100</concept_significance>
%</ccs2012>
%\end{CCSXML}

%\ccsdesc[500]{Computer systems organization~Embedded systems}


\keywords{cognitive computing, cloud computing, edge computing, mobile cloud computing, service migration}

\setcopyright{acmcopyright}
\acmJournal{TOIT}
%\acmYear{2018} \acmVolume{1} \acmNumber{1} \acmArticle{1} \acmMonth{1}
\acmPrice{15.00}\acmDOI{10.1145/3239565}

\maketitle

%\renewcommand{\shortauthors}{G. Zhou et al.}

\section{Introduction}

The massive proliferation of personal computing devices is opening up new human-centered designs that blur the boundaries between humans and machines \cite{Sarros2018edge}.
Now, the frontier for research on  data management is related to the so-called edge computation and communication, consisting of an architecture of one or more collaborative multitude(s) of computing nodes that are positioned among edge networks with the access of cloud-based services. Such a mediating level is responsible for carrying out a substantial amount of data storage and processing to reduce the retrieval time and have more control over the data with respect to cloud-based services, and to consume fewer resources and energy to reduce the workload \cite{wei2016, eciot}.

The edge computing paradigm has multiple advantages. Firstly, the edge node can reduce the traffic load of backhaul by providing a certain amount computing capability, which is significant for applications such as an online games that need to transmit 60 or even 120 frames per second. As an alternative solution, the server only sends parameters such as character position, timestamp and attribute changes (some common data), and leaves the edge node to compute and render the visual image. Secondly, as a result of the large number of edge nodes deployed in 5G and the big-data analysis based on user preference, the popular contents can be acquired in advance in the interconnecting edge devices, which are only one hop away from the user.

However, edge computing is also faced with many challenges. Firstly, the operation and processing capabilities of an edge device are limited and can fail to meet the demands on real-time service, data optimization, and application intelligence. Secondly, the intelligence of most typical edge-computing services is only embodied in the artificial intelligence (AI)-enabled data storage and processing on the edge. However, the intelligence is missing from the aspects of behavior feedback, automatic networking, load balance, and data-driven network optimization.

%\subsection{Cognitive Computing}
Cognitive computing originates from cognitive scientific theory. Now, it makes machines achieve ``brain-like'' cognitive intelligence through an interactive cognition loop with machine, cyber space and humans.
Compared to big-data analytics, it possesses the following features: 1) analysis of the existing data and information in  cyber space, to improve the intelligence of the machine; 2) the machine reinterprets and explains the information in the existing cyber space, and accordingly generates new information; humans also participate in this process; 3) the cognition of the machine to the human provides a more intelligent cognitive service.
Its enabling paradigms (e.g., agent-based computing) had been researched and the related concrete applications based on cognitive Internet of Things (IoT) platforms and frameworks have been studied in \cite{agent16, gian15, gian14}.

Nevertheless, the cognitive computing application mainly depends on the machine-learning model trained on the cloud, while the real-time inference requests are made by end edge devices, which so far have been the most common deployment mode of the cognitive service. The existing problem of such a mode is the large latency in the network operation and the service delivery. However, if the cognitive service is deployed on the network edge, the latency of the network response to the user request will be greatly reduced, so the research of edge deployment for training and inference machine is rapidly increasing.

Therefore, considering both the advantages and disadvantages of edge computing and cognitive computing, a new computing paradigm called Edge Cognitive Computing (ECC) is proposed, which combines edge computing and cognitive computing.
Such a new architecture integrates communication, computation, storage and application on edge networks; it can achieve data and resource cognition by cognitive computing. Moreover, it can provide personalized services nearby, enabling the network to have a deeper, human-centered cognitive intelligence.

The main contributions of this paper are as follows:
\begin{itemize}
  \item We propose a new ECC architecture that deploys cognitive computing at the edge of the network to provide dynamic and elastic storage and computing services. In addition, the design issues of how to fuse these key technologies of cognitive computing and edge computing are illustrated in detail.
  \item We propose an ECC-based dynamic cognitive service-migration mechanism that considers both the elastic allocation of the cognitive computing services and the user mobility, to provide a mobility-aware dynamic service-adjustment scheme.
  \item We develop an ECC-based test platform for dynamic service migration and evaluate it by means of several experiments, with the results showing that the proposed ECC can provide dynamic services according to different user demands.
\end{itemize}

The remainder of the paper is organized as follows. Section II introduces the related work. Section III presents the proposed architecture for edge cognitive computing in detail. Section IV stipulates the related design issues. Section V introduces the dynamic cognitive service migration mechanism. Section VI demonstrates the ECC test platform for dynamic service migration. Finally, Section VII concludes the paper and reports some open issues.

%%--------------------------------------------
\section{Related Work}
%%--------------------------------------------

The need for on-demand state-of-the-art services (smart sensing, e-healthcare, smart transportation, etc.) and the latency issues that affect the overall Quality of Service (QoS) for various applications have paved the way to the powerful paradigm of edge computing.
%The edge-computing architecture is shown in Fig. \ref{fig003}.

%\begin{figure}
%\centering
%\includegraphics[width=\columnwidth]{Fig03.eps}
%\caption{Architecture of edge computing.}
%\label{fig003}
%\end{figure}

There is a lot of research on edge computing with respect to energy efficiency and latency. The cooperation and interplay among cloud and edge devices can help to reduce energy consumption in addition to maintaining the QoS for various applications.
However, a large number of migrations among edge devices and cloud servers leads to congestion in the underlying networks. Hence, to handle this problem, \cite{Edge18IIoT} presented an SDN-based edge-cloud interplay to handle the streaming of big data in the Industrial IoT environment.
Edge computing is expected to support not only the ever-growing number of users and devices, but also a diverse set of new applications and services. The work in \cite{Sarros2018edge} introduced a system that can pervasively operate in any networking environment and allows for the development of innovative applications by pushing services near to the edge of the network.

In terms of security and privacy in edge computing,
there is an increasing realisation that edge devices, which are closer to the user, can play an important part in supporting latency and privacy-sensitive applications \cite{edge17rana, iot17singh}.
Therefore, the security challenges relate to the protection of device data, such that an unauthorized person cannot access the data, providing secure data sharing between the device and the edge cloud, and safe data storage on the edge cloud \cite{secu17rana,zhou17,zhou18}.

However, the above researches on edge computing mostly focused on solving the communication problems by leveraging computing and storage, like how to reduce the network load, improve the network efficiency and reduce the transmission delay. In addition, these works did not consider how to solve the personalization of actual AI applications and how to provide elastic storage and computing services.

Applying cognitive computing in various applications for smart cities has been widely researched, for example, authors in \cite{soft18} studied the role of intelligence algorithms such as machine learning and data analytics within the framework of smart-city applications such as smart transportation, smart parking and smart environment, to address the challenges of big data. \cite{cog18city} also explored how deep reinforcement learning and its shift towards semi-supervision can handle the cognitive side of smart-city services.
The work in \cite{deep2018a} indicated that the application of deep networks has already been successful in big-data areas, and fog-to-things computing can be the ultimate beneficiary of the approach for attack detection because the massive amount of data produced by IoT devices enables deep models to learn better than shallow algorithms.
In summary, most of the current researches on cognitive computing has focused on the design of algorithms. However, if cognitive computing wants to be applied and deployed on a large scale, it is necessary not only to solve the problems of how to compute, but also to solve what to compute and where to compute, which needs to deploy cognitive computing at the edge of the network.

There are some researches on applying cognitive computing to edge computing in \cite{learnEdge}. The authors first introduced deep learning for IoTs into the edge-computing environment. They also designed a novel offloading strategy to optimize the performance of IoT deep-learning applications with edge computing since the existing edge nodes have limited processing capability.
The works in \cite{deep18veh} and \cite{sdn17deep} proposed a novel deep-reinforcement-learning approach to solve the resource allocation problems in terms of networking, caching and computing in edge computing. However, the above researches did not apply the cognition for applications to guide network-resource optimization, but only considered the resource allocation using some intelligent algorithms, which cannot provide elastic cognitive computing services.

%%--------------------------------------------
\section{The Proposed ECC Architecture and Design Issues}
\begin{figure*}
\centering
\includegraphics[width=0.94\columnwidth]{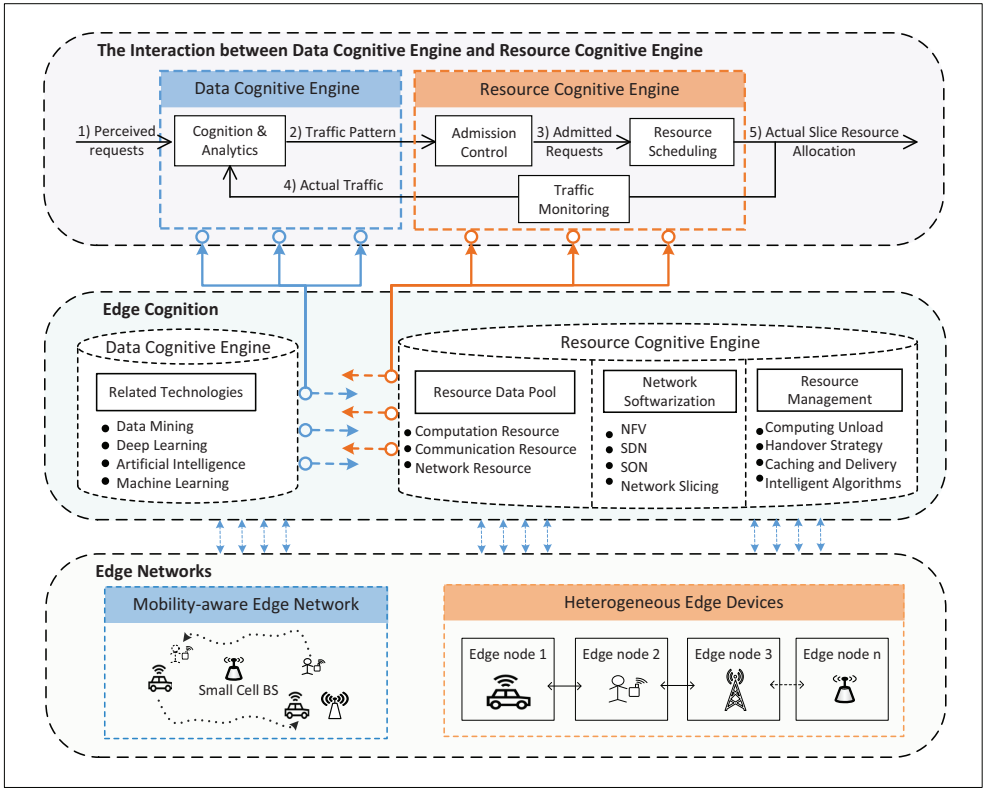}
\caption{The edge cognitive computing architecture.}
\label{fig002}
\end{figure*}

%%--------------------------------------------
%The ECC is aimed at providing more intelligent and dynamic cognitive services by edge devices, utilizing the edge resources efficiently and reducing the cognitive resource consumption.
The proposed ECC architecture mainly consists of two components: \emph{the edge network} and \emph{the edge cognition} as shown in Fig. \ref{fig002}. The edge network mainly provides the access and resource management of various edge devices. The edge cognition mainly relates to the cognition to edge data, involving service data and network and computing resource data. The edge cognition is mainly composed of two core parts, data cognitive engine and resource cognitive engine. The interaction between data cognitive engine and resource cognitive engine is the key design issue which is also shown at the top of Fig.~\ref{fig002}.

%The human-centric cognitive cycle of cognitive computing can properly realize intelligence in terms of behavior feedback, automatic networking, load balancing and data-driven network optimization in edge networks. In addition, it can better satisfy the demands of real-time business, data optimization and application intelligence in the edge environment.

In the architecture, the data cognitive engine mainly relies on cognitive computing technologies while the resource cognitive engine mainly uses the related technologies of edge computing.
By combining key technologies in cognitive computing (i.e., big-data analysis, machine learning, deep learning) with those in edge computing (i.e., computing offload and migration, mobility management, intelligent proactive caching, resource cooperation management), ECC can better solve the problem of communication bandwidth and delay by the fusion of computing, communication,and storage, and thus improving the network intelligence.
Below we will introduce the ECC architecture in details from three aspects: resource cognitive engine, data cognitive engine and the interaction between them.

\subsection{Resource Cognitive Engine}
This engine can learn the characteristics of edge cloud computing resources, environmental communication resources, and network resources by cognition, and feedback the integrated resource data to the data cognitive engine in real time. At the same time, it can accept the analysis result of the data cognitive engine and realize the real-time dynamic optimization and allocation of resources. As shown in Fig. \ref{fig002}, it mainly includes the resource data pool, network software technologies and resource management technologies. More specifically, the function of this engine is to:
\begin{enumerate}
\item \textbf{Resource Data Pool:} Realize the massive, heterogeneous and real-time connection between terminals (such as smart clothing, intelligent robot, intelligent traffic car, and other access devices), ensure the security, reliability and interoperability of connection, and constitute the resource data pool (computing resources, communication resources and network resources) as a basic architecture for data transmission.
\item \textbf{Network Softwarization:} Utilize the network software technologies involving network function virtualization (NFV), software-defined network (SDN), self-organized network (SON), and network slicing to realize high reliability and flexibility, ultra-low latency, and extendibility of the edge cognitive system.
\item \textbf{Resource Management:} Utilize the resource-management technologies involving computing unload, handover strategy, caching and delivery, and intelligent algorithms to build a cognitive engine with resource optimization and energy saving to enhance QoE and meet the different demands of various heterogeneous applications.
\end{enumerate}

%\subsection{Key mechanisms in the resource cognitive engine}
The key mechanisms involved in the resource cognitive engine include network slicing, computing unloading, caching and delivery, etc.
Using the virtualization technology, network slicing virtualizes the physical infrastructure of 5G into multiple virtualized network slices that are mutually independent and parallel to realize the arrangement and management of the corresponding network resources.
The computing unloading is responsible for the consideration of the computing tasks' assignment problem, aiming to rationally allocate the computing resources on the edge cloud and remote cloud, thus to complete the computing tasks through cooperation.
By caching and delivering, the predicted contents are placed on the edge in advance, and thus the low latency and load reduction of the core network are achieved.
SDN/NFV can reduce the deployment costs and improve the efficiency of the network control through the virtualization of the network resources.

\subsection{Data Cognitive Engine}
This engine deals with the real-time data flow in the network environment, introduces the data analysis and automatic service processing capabilities to the edge network, realizes cognition to the service data and the resource data by using various cognitive computing methods, including data mining, machine learning, deep learning, and artificial intelligence as shown in Fig. \ref{fig002}. The main data sources are:

\begin{enumerate}
\item Collect the external data from the data source in the application environment, such as physical signs and real-time disease risk level under the cognitive health surveillance, or real-time behavior information on the mobile user;
\item Collect dynamically the internal data on computing resources, communication resources and network resources of the edge cloud, such as network type, service data flow, communication quality, and other dynamic environmental parameters.
\end{enumerate}

%\subsection{Key issues for the data cognitive engine}

The key point of the intelligent enhancement of data cognition engine is that multi-dimensional data (including external data related to the user and the service, and the internal data in the resource network environment) are adopted in cognitive computing technology, which is not the case in the traditional data-analysis methods.
The data cognitive engine conducts an analysis of the existing data and information (e.g., using deep convolutional network (DCNN) for facial emotion recognition and using hidden markov model (HMM) for user mobility prediction). It then feeds them back to the resource cognitive engine, after which the resource cognitive engine conducts a reinterpretation and analysis of the information to generate new information, which may be further utilized by the data cognitive engine.
For instance, in health monitoring, after the monitoring and analysis of the physical health of a smart-cloth-worn user using cognitive computing, a health-risk level of that user will be obtained; then the resource allocation in the whole edge-computing network will be comprehensively adjusted to the risk level of each user, i.e., the data are utilized for the second time and serve resource allocation and network optimization in turn to form a closed-loop system for cognitive intelligence.

%%--------------------------------------------
%\section{Design Issues for Edge Cognitive Computing}
%%--------------------------------------------

\subsection{The Interaction between Data Cognitive Engine and Resource Cognitive Engine}

The key design issue of the ECC is the interaction between data cognitive engine and resource cognitive engine. In the edge cognitive computing, we put forwards the design idea of realizing the closed-loop optimization with the double cognitive engine to optimize the network resource management technology such as network slice. Here we take cognitive network slicing as an example to illustrate how to fuse the related technologies in cognitive computing and edge computing.

As shown in Fig. \ref{fig002}, data cognitive engine first perceived many requests. The request types of the network-slice service differ from one to another according to different demands (latency, reliability and flexibility) of different cognitive applications. Then data cognitive engine will conduct the fusion cognitive analysis of the heterogeneous data based on current resource distribution situation and real-time requests of the tenant with methods of machine learning and deep learning. Next the data cognitive engine will report the analyzed dynamic traffic pattern to the resource cognitive engine. In the resource cognitive engine there is a joint optimization of the comprehensive benefits and the resource efficiency. Firstly, conduct admission control to perceived requests, then conduct the dynamic resource scheduling and distribution based on the cognition of network resources, and feed the scheduling results back to the data cognitive engine, to realize the cognition of the network-slice resources.

%The data cognitive engine and resource cognitive engine refer to the computation unloading issue in the interactive process. Moreover, the computation unloading decision and execution are influenced by many factors, such as the mobility, the number of users and the task type. In the cognition edge computing, similar to the network-slice cognition, the user behavior prediction provided in combination with the data cognitive engine can realize the on-demand resource distribution and intelligent mobility perception prediction better, reduce the computing energy consumption, and improve the success rate of the unloading.

%%--------------------------------------------
\section{Dynamic Cognitive Service Migration Mechanism}
%%--------------------------------------------

Under the ECC architecture, due to the mobility of the user, the heterogeneity of the edge device and the dynamics of the network resources (such as the available storage, the computing resources and the network bandwidth), we should offer the elastic cognitive service, i.e., offering the service in accordance with the personalized demands of the user. The amount of computation consumed by cognitive computing is particularly large, so the computing resources are required to be more elastic and flexible if deploying the cognitive computing on the edge.
The ECC proposed in this paper is different from that proposed by those in related work. The ECC mainly focuses on applications related to the artificial intelligence in the IoT, such as automatic pilot, virtual reality, smart clothing, Industry 4.0, emotion recognition, etc. In contrast to traditional content retrieval and mobile computing issues, such applications are often more personalized, so the computing resources are required to be more elastic and flexible.

To describe the proposed ECC architecture better, we implemented the Dynamic Cognitive Service Migration Mechanism. Because the device bearing the computing is varying, a service migration mechanism is needed.
In our ECC-based dynamic service migration mechanism, to reduce the latency, the workload should better be finished in the nearest edge device that has enough computation capability at the edge of the network.
Thus, according to the user behavior prediction, some contents needed for the service or some jobs for the task are migrated in advance, or the low-resolution work is firstly migrated to the position to be moved. After the user's pass-by, the service resolution is promoted on that device so offering the elastic service.

\subsection{Service Resolution}

To better explain the elastic service provided by the ECC, we define a new metric called service resolution to evaluate the user QoE. In view of the different applications, the service resolution has different definitions. For example, the emotion detection depends on the accuracy rate and the latency of the emotion recognition, while both are mutually contradictory. A higher accuracy rate needs more computing resources, with higher latency. However, when the user is insensitive to the accuracy rate and pays more attention to the interactive experience, we can provide a low resolution without influencing the user QoE. For the application of video streaming, the service resolution depends more on the resolution of the video streaming acquired by the user.
Table \ref{tab.resolution} lists the service resolution of the two different applications. The emotion detection deems the accuracy rate as a metric, and the video streaming deems the resolution as a metric, respectively offering three services to meet the QoE under different demands of the user, i.e., offering the elastic service and enhancing the user experience.

\begin{table*}
\renewcommand{\arraystretch}{1.1}
\caption{Service Resolution for Different Applications.}
\label{tab.resolution} \centering
\begin{center}
\begin{tabular}{|c|c|c|c|c|}
\hline
\multirow{2}{*}{\textbf{Applications}}	& \multicolumn{3}{c|}{\textbf{Service Resolution}} & \multirow{2}{*}{\textbf{Main Metric}}\\
\cline{2-4}
    & Low  & Middle  & High & \\
\hline
Emotion Detection & 66.3  & 73.6  & 79.1  & Accuracy (\%) \\
\hline
Video Streaming & 800$\times$600  & 1280$\times$1024  & 1920$\times$1080  & Video resolution (pixels) \\
\hline
\end{tabular}
\end{center}
\end{table*}

We will explain how to offer the elastic cognitive computing service from the perspective of the two applications, as follows.

%\subsubsection{Emotion Detection}

\begin{figure*}
\centering
\subfigure[]
{\includegraphics[width=0.8\columnwidth]{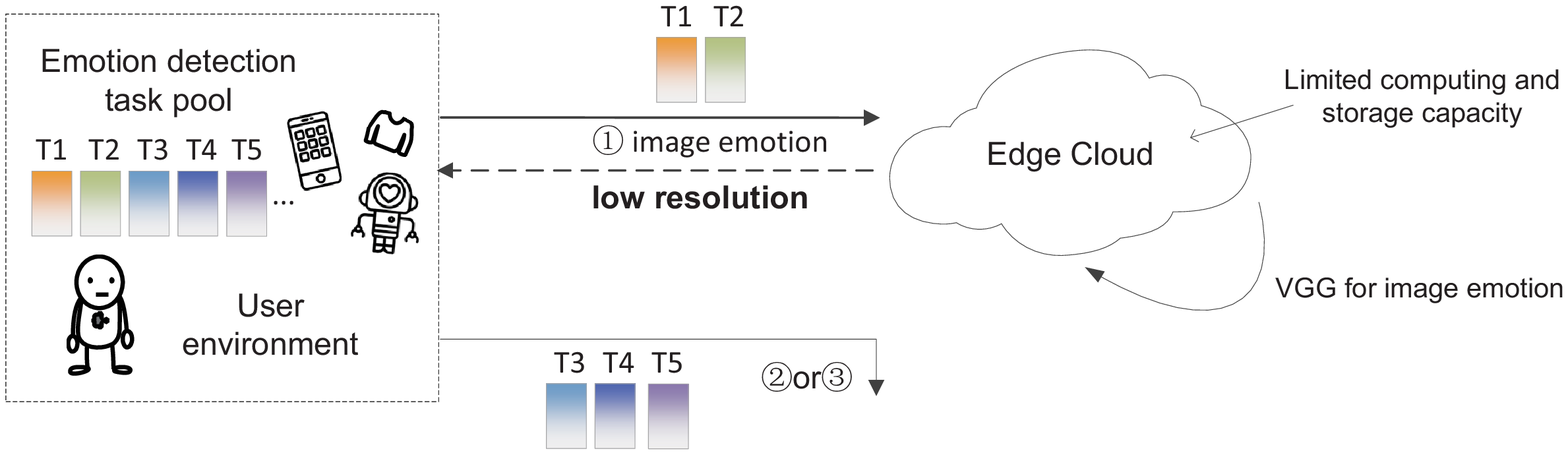}
\label{emotiona}
}
\subfigure[]
{\includegraphics[width=0.8\columnwidth]{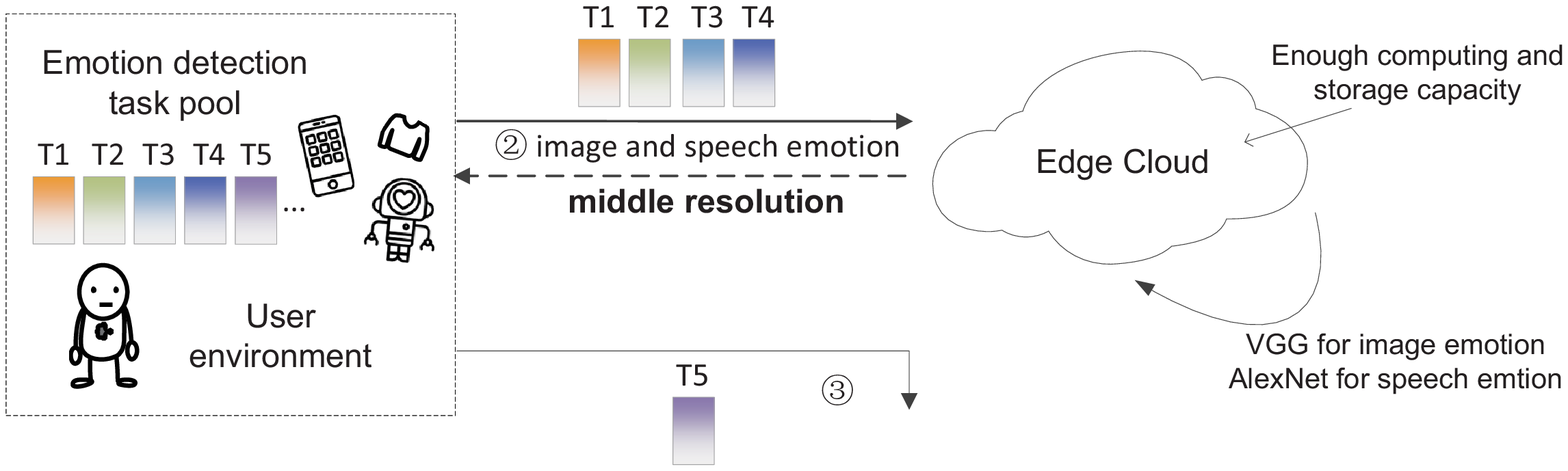}
\label{emotionb}
}
\subfigure[]
{\includegraphics[width=0.8\columnwidth]{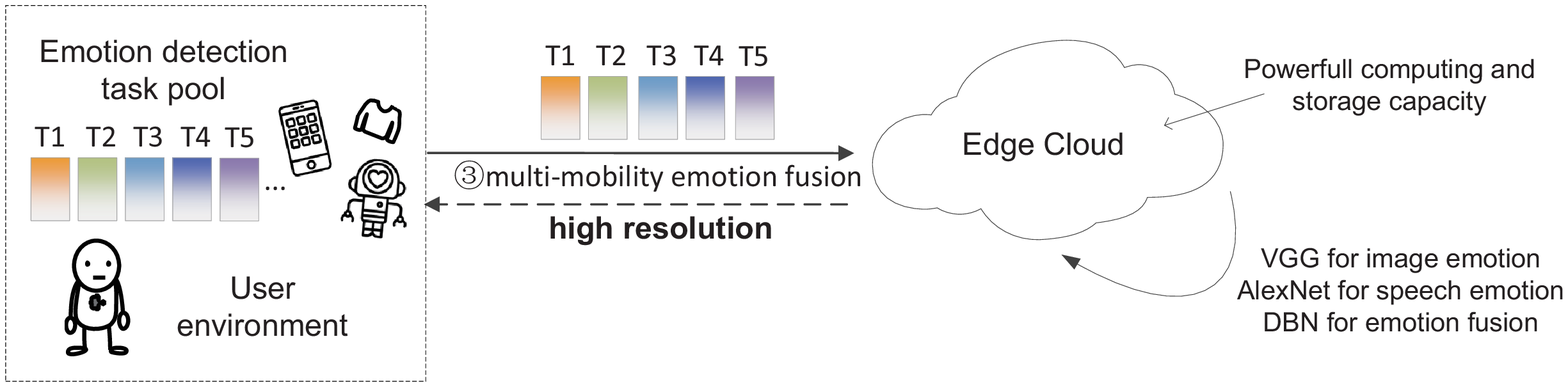}
\label{emotionc}
}
\caption{Service resolution for emotion detection: (a) low resolution, (b) middle resolution, (c) high resolution.}
\label{emotion}
\end{figure*}

\textbf{1) Emotion Detection:} As shown in Fig. \ref{emotion}, we provide three service resolutions for emotion detection: low resolution, medium resolution and high resolution. In the case of limited computing resources, we provide low resolution, i.e., only conducting the facial emotion recognition and using the deep neural network VGG. For the medium resolution, we analyze the facial expression (VGG \cite{vgg14}) and speech emotion (AlexNet \cite{alex17}) simultaneously, and carry out the simple decision fusion. For high resolution, we use the strong computing resources, provide the multi-modal emotion recognition algorithm, and use the deep network i.e., deep belief network (DBN) for the decision fusion. For these three service resolutions, the computing resources consumed are increased gradually, and the accuracy rate of the emotion recognition provided is higher.

The user of emotion detection is always a mobile user, so the dynamic change of the mobile computing resources is one of factors influencing the user QoE. In addition, when the user is moving, the network status is changed, but the emotion recognition is required to maintain ultra-high reliability during the communication process. Thus, it is necessary to adopt the elastic computing mode to solve this problem. This application in need of multiple computing decisions was not considered in  previous research.
It is a mutual contradiction of ensuring the accuracy rate and the latency of emotion recognition at the same time, and a higher accuracy rate needs more computing resources, with higher latency, as shown in Table \ref{tab.acc}. However, when the user is insensitive to the accuracy rate and pays more attention to the interactive experience, we can provide low resolution without influencing the user QoE.

\begin{table}
\renewcommand{\arraystretch}{1.1}
\caption{Accuracy and Latency of different service resolutions for emotion detection.}
\label{tab.acc} \centering
\begin{center}
\begin{tabular}{|l|l|l|}
\hline
\textbf{Algorithms}	& \textbf{Accuracy (\%)}	 & \textbf{Latency (ms)} \\
\hline
VGG & 66.3 & 103.0 \\
AlexNet + VGG & 73.6 & 188.4 \\
AlexNet + VGG + DBN & 79.1  & 265.3 \\
\hline
\end{tabular}
\end{center}
\end{table}

%\subsubsection{Video Streaming}

\textbf{2) Video Streaming:} Similar to emotion detection, we provide three service resolutions for a video streaming application, i.e., in consideration of different user demands, user mobility and a dynamic network environment simultaneously, we provide the video decoding with different resolutions, respectively, and decompose the video decoding task into different resolution tasks in a similar way. When the user is moving, the edge device node better judges whether to conduct the task migration and which resolution of the task migration is conducted according to the user mobility behavior. For example, when the user moves to the other edge node without determining a long-term stay or a short-term stay, the video decoding task with low resolution can be firstly migrated. In the case of a long-term stay of the user, the high-resolution service can be offered, for avoidance of the untimely migration and resource waste. In addition to considering the user mobility, the migration costs should be considered. The low-resolution service has the lowest migration cost, and the high-resolution service has the highest migration cost.

\subsection{Dynamic Service Migration Mechanism}

When and how to conduct migration are the two major concerns in dynamic service migration mechanisms. Most migration mechanisms decide when to migrate only relying on network condition, few of them take the user behavior into account \cite{mig17,mig14}. However, deciding when to migrate according to user behavior and mobility has large influence on improving user experience and resource utilization.

\begin{figure}
\centering
\includegraphics[width=0.7\columnwidth]{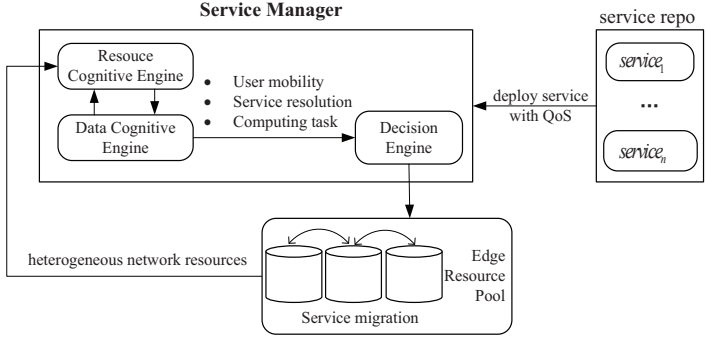}
\caption{Service deployment architecture for ECC.}
\label{service}
\end{figure}

As shown in Fig. \ref{service}, the Service Manager implements all the functionalities that an edge node needs to deploy its services. It includes a service repository (service repo) where services ($service_1,...,service_n$) to be provided are stored, e.g. dockerized compressed images or the emotion recognition models.
The Decision Engine is responsible for deciding which services to deploy. In Fig. \ref{service},
the resource cognitive engine manages the computing and network resources of the heterogeneous edge device, and cognizes the user mobility, user demands for service resolution and resource demands for computing tasks in combination with the data cognitive engine. The Decision Engine makes the decision in accordance with the information and migration strategy (based on Q-learning, see below), and accordingly offers the dynamic and elastic cognitive services.

The service providers (SPs), i.e., the edge nodes, manage the virtual networks and let $\mathcal{M} = \{1,...,M\}$ be the set of SPs.
Let $t \in \{0,1,2,...,N\}$ denote the time instant of service request.

%另一种 Task 定义
We assume that the edge device has $n$ services that need to be migrated and the set of tasks is denoted as $\mathcal{T}=\{T_1,T_2,...,T_n\}$. For the migration task $T_i$, $T_i=\{\omega_i, s_i, o_i\}$, where $\omega_i$ is the amount of computing resource required for the task $T_i$, i.e., the total number of CPU cycles needed to complete the task, and $s_i$ is the data size of the computation task $T_i$, i.e., the amount of data content to be delivered to the other edge node, specifically, in this work, it stands for the size of the video content or the storage resource consumed by the emotion detection (e.g., the processing code and parameter(s)). Finally, $o_i$ represents the data size of the task result. For instance, in the video decoding case, $\omega_i$ is the computing resource needed for the video decoding, $s_i$ is the video data size, and $o_i$ is the data size of the decoded video. After the computation, the Service Provider $m$ sends the transcoded video content back to the user.

\textbf{Migration Cost:} The traffic volume of migrating a virtual server usually cannot be neglected due to the large size of the server states. The migration cost of a virtual server depends upon the size of the server  as well as the bandwidth available on the migration path. For example,
%对于 emotion detection 服务来说，迁移代价取决于 the emotion recognition models，对于video streaming 服务来说，迁移代价取决于 the data size of the decoded video。而 higher service resolution，其迁移代价也更高。
for the emotion detection service, the migration cost depends on the emotion recognition models. For a video-streaming service, the migration cost depends on the data size of the decoded video. A higher service resolution has a higher migration cost.

\textbf{Migration Goal:} Minimize the service costs, and in the meantime, improve the QoE by providing different service resolutions based on
%用户需求、用户移动性和动态的网络资源. 对于服务请求$x_t$, 我们定义在某种迁移策略$\pi$下的Score 为$Score(x_t, \pi)$，代价为$Cost(x_t, \pi)$, 那么优化目标可定义如下：
user demands, user mobility and dynamic network resources. For $x_t$, a service request at time $t$, we define the score (the metric for user acquired experience) under some migration strategy $\pi$ as $Score(x_t, \pi)$ and the cost of $Cost(x_t, \pi)$, so the optimization objective can be defined as follows.
\begin{equation}\label{eq.01}
    max \ F(x, \pi)= \sum_{t\in {0,1,...N}}Score(x_t, \pi)- Cost(x_t, \pi)
\end{equation}

%其中 $Score(x_t, \pi) = \frac{R(x_t)-E(x_t)}{Delay(x_t)}$, $R(x_t)$ 为服务请求所获取的服务类型，即service resolution, 我们设置值为(0,1,2)分别对应于low, middle and high service resolution. $E(x_t)$ 为服务请求的期望。$Delay(x_t)$ 为获取服务的时间，与$\omega_i$和$o_i$ 等有关. $Cost(x_t, \pi)$ 与$s_i$有关.

Where, $Score(x_t, \pi) = \frac{R(x_t)-E(x_t)}{Delay(x_t, \pi)}$, $R(x_t)$ is the service type acquired by the service request, i.e., the service resolution. We set the value of (0, 1, 2), respectively corresponding to low, medium and high service resolution. $E(x_t)$ is the expectation of a service request. $Delay(x_t, \pi)$ is the time of service acquisition under strategy $\pi$, relevant to $\omega_i$ and $o_i$. $Cost(x_t, \pi)$ is relevant to $s_i$.

%通过the definition of $Score(x_t, \pi)$ 可以看出，当时延一定，用户对服务需求一定时，所提供的service resolution 越高，获得越高的用户体验；当提供同样的service resolution 时，用户期望越高，得分越小，$R(x_t)-E(x_t)$ 能很好地反映用户所获取的服务与用户期望的关系，这也意味了当用户对服务质量要求不高时，我们可以提供 low resolution 的服务，这样可以降低能耗，同时不影响用户QoS。 当用户所获取的服务与用户期望一定时，获取服务的delay 越高，Score越低。

From the definition of $Score(x_t, \pi)$, it is observed that, in the case of a definite latency and a definite service demand of the user, the higher the service resolution is provided, and the higher the user experience is gained. While providing the same service resolution, the higher the user expectation is, the lower the score is. $R(x_t)-E(x_t)$ can well reflect the relationship between the service acquired by the user and the user expectation. This means that we can provide the low-resolution service if the quality of the service requirements of the user is not high, so as to reduce the energy consumption, without influencing the user QoE. When the service acquired by the user and the user expectation are definite, the higher the delay of service acquired is, the lower the score is.

\textbf{Optimal Problem Formulation:} Our problem can be described as a reinforcement learning scenario. The objective is to find an agent that makes the optimal migration policy for each service request. The optimal migration policy denoted by $\pi^*$ can maximize the system reward given by:
\begin{equation}\label{eq.02}
  \pi^* = \operatorname*{argmax} \limits_{\pi}\sum_{x\in \mathcal{X}}F(x, \pi)
\end{equation}

Let $S_i$ denote the state of environment at time $i$, defined by the locations of the $n$ services at that time. For a sequence of batch requests $\mathcal{X} = \{x_1, x_2, ..., x_N\}$, the goal of the service migration is to determine $S_1,S_2,...,S_N$ to maximize the system reward defined by Eq. (\ref{eq.02}).

Q-learning is one of the most popular Reinforcement Learning \cite{RL} (RL) methods that is applied in many research areas.
The general procedure of the Q-learning algorithm is shown as the Algorithm \ref{alg:qlearning}.

We define the reward after the action $a$ taken on $S_t$ as:
\begin{equation}\label{eq.03}
  R_{t+1}^a = Score(S_{t+1})-(Score(S_t) + Cost(S_t,S_{t+1}))
\end{equation}

Similarly, we can also construct a matrix $Q$ to memorize the experience that the agent has gained from the environment. The Q-value of the state-action pair, $Q(S_t, a)$, represents the expected total benefits caused by action $a$ taken in state $S_t$. The solution is to exploit from the initial state to a final optimal state through updating accordingly by Algorithm \ref{alg:qlearning}. In each iteration of the algorithm, the agent observes the current state $S$ and takes action $a$ to move to the next state $S'$ by receiving an immediate reward $R_{t+1}$, which is used to update the $Q(s,a)$ by following Eq. (\ref{eq.04}), and then begins the next iteration.
\begin{equation}\label{eq.04}
  Q(S_t, A_t) \gets Q(S_t, A_t) + \alpha(R_{t+1} + \gamma \max \limits_aQ(S_{t+1}, a) - Q(S_t, A_t))
\end{equation}

The $\alpha$ means learning rate which determines how much the new information overwrites the old one. The discount factor $\gamma$ gives more weight to the most recent reward than others in the future.

\begin{algorithm}[htb]
\caption{Q-learning algorithm.}
\label{alg:qlearning}
Initialization $Q(s,a)$ \\
Repeat (for each episode):\\
\hspace*{0.15in} Initial state S \\
\hspace*{0.15in} Repeat (for each step in episode):\\
\hspace*{0.3in} Use some policy such as ($\epsilon$ - greedy), and select an action for execution based on the state $S$ \\
\hspace*{0.3in} After executing the action, observe reward and new state $S'$ \\
\hspace*{0.3in} $Q(S_t, A_t) \gets Q(S_t, A_t) + \alpha(R_{t+1} + \gamma \max \limits_aQ(S_{t+1}, a) - Q(S_t, A_t))$ \\
\hspace*{0.3in} $S \gets S'$ \\
\hspace*{0.15in} End
\end{algorithm}

%%--------------------------------------------
\section{Testbed and Performance Evaluation}
%%--------------------------------------------

To verify the proposed architecture, an ECC test platform was set up, and a performance evaluation of the dynamic service migration mechanism was conducted in the experimental testbed for the user mobility.

\begin{figure*}
\centering
\includegraphics[width=0.8\columnwidth]{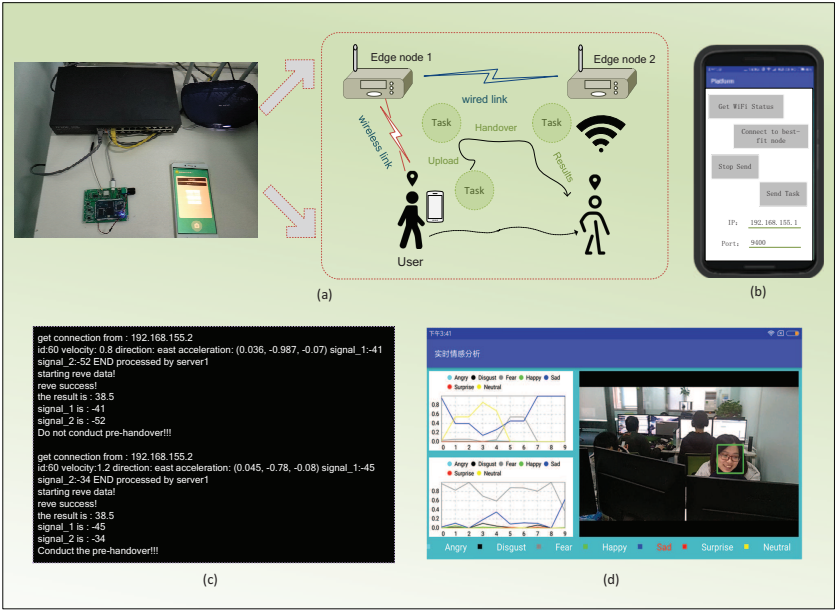}
\caption{Experimental platform: (a) hardware platform, (b) interface of mobile application, (c) software interface of the edge node, (d) UI of emotion detection application.}
\label{fig006}
\end{figure*}

%\subsection{Experimental Platform}
To create the ECC environment, we used several edge-computing nodes that realize the functions of emotion detection and video streaming as shown in Fig \ref{fig006}(a). The self-designed hardware is with core processor of 4 x ARM Cortex-A33 and the algorithm execution in edge device needs deploying TensorFlow environment. We also used an Android phone as a user mobile device, and designed the Android application program as shown in Fig \ref{fig006}(b), and realized the signal monitoring of the edge-computing node, task uploading, result downloading, and service migration.
Fig \ref{fig006}(c) illustrates the software interface running on Windows. Fig \ref{fig006}(d) shows the UI of the emotion detection application.

%\subsection{Experimental Settings}

In the experimental setup, we use four edge nodes and two servers, i.e., $m=4, n=2$.
For the performance comparison, two schemes are compared with the proposed ECC-based scheme: 1) No migration scheme: service migration is not considered; 2) Nearest migration scheme: if needed, service will be migrated to close access point.

Table \ref{tab.04} lists the values of important parameters considered in the experiments. The task load in the high-resolution migration was 256 MB, the task load in the middle-resolution migration was compressed to 128 MB, and the task load in the low-resolution migration was compressed to 64 MB. The transmission bandwidth between the edge nodes was 5 Mbps.

\begin{table*}
\renewcommand{\arraystretch}{1.1}
\caption{Experimental parameters.}
\label{tab.04} \centering
\begin{center}
\begin{tabular}{lll}
\hline
\hline
\textbf{Parameter}	& \textbf{Value}	 & \textbf{Description} \\
\hline
$B_{i,j}$ & 5 Mbps	 & The bandwidth between edge node $SP_i$ and $SP_j$. \\
\hline
$Q_{T_i}$ & 100 Mcycles  & The required number of CPU cycles to complete task $T_i$. \\
\hline
$O_{T_i}$ & 1 Mbits & The content size for task $T_i$. \\
\hline
$\alpha$ & 0.01 & The learning rate of algorithm. \\
\hline
$\gamma$ & 0.8 & The discount factor giving more weight to the near future. \\
\hline
\hline
\end{tabular}
\end{center}
\end{table*}

%\subsection{Experimental Results}

\begin{figure}
\centering
\includegraphics[width=0.6\columnwidth]{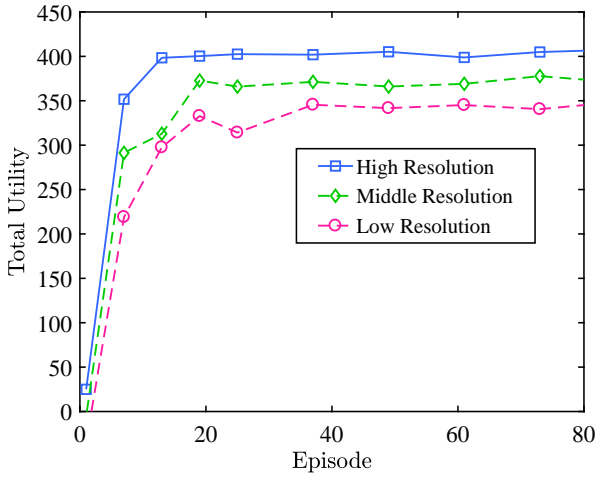}
\caption{Convergence performance with difference service resolutions.}
\label{fig007}
\end{figure}

\begin{figure}
\centering
\includegraphics[width=0.6\columnwidth]{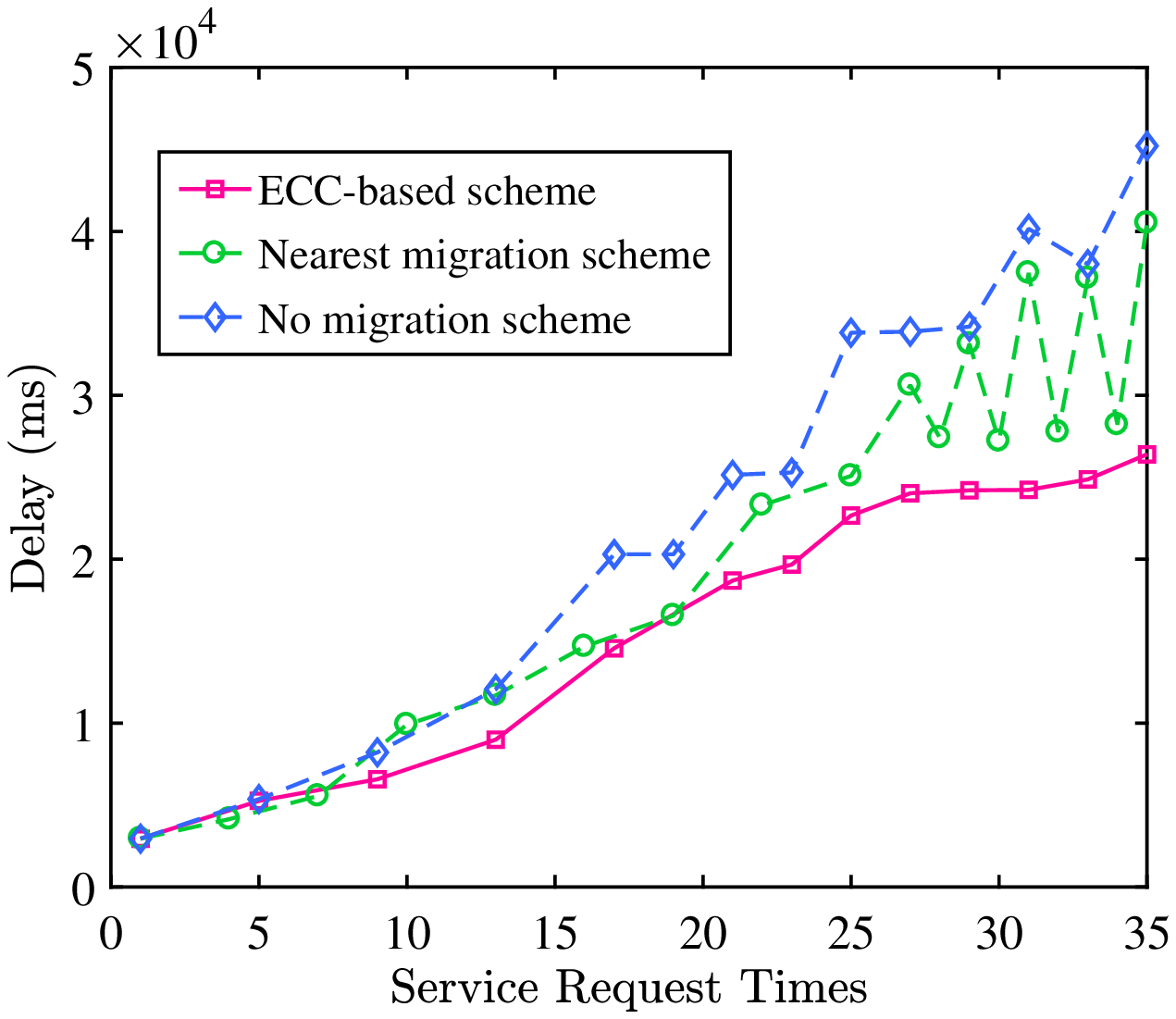}
\caption{Delay comparison of different migration schemes with request times.}
\label{fig008}
\end{figure}

Fig. \ref{fig007} and \ref{fig008} plot the experimental results for the performance analysis.
Fig. \ref{fig007} shows the convergence performance of different scenarios in the proposed scheme using the deep reinforcement learning algorithm. From Fig. \ref{fig007} we can see that the total utility (the cumulative rewards, i.e., the object function $F$ defined in Eq. \ref{eq.01}) of the different scenarios in the proposed scheme is very low at the beginning of the learning process. With the increase in the number of episodes, the total utility increases until it reaches a relatively stable value, which is around 400 in the scenario that provides high resolution.
%We can also observe that 刚开始不同service resolution获得的收益差别不大，到稳定阶段，high resolution 获得reward 更大. Therefore, 在预测到用户将移动到其他边缘接入点时，可以先将low resolution 的服务迁移过去，再稳定阶段提供high resolution 的服务.
We can also observe that the rewards of different resolution services are almost the same at the beginning, and at a stable stage high resolution obtains the highest reward, while low resolution obtains the lowest reward. Therefore, the low-resolution service could be first migrated to the corresponding edge node when the user moves to other edge access point, then the high-resolution service should be provided at a stable stage.

%Fig. \ref{fig008} 显示了边缘节点提供emotion detection服务所需的响应时间的曲线变化。通常来说响应时间与请求并发数成正比，which means that 并发请求数越多，单次请求的平均响应时间越长。我们比较了不同 schemes 下的时延，from Fig. \ref{fig008}, we can obtain that 三个延时时间均随并发请求数的增加而增多，our proposed RL-based scheme under ECC architecture has better performance with lowest latency. This is because that 我们对用户移动性进行了预测，能更好地提前将服务迁移到离用户请求接入点最近的地方，而一般的迁移方法则是当用户移动到另一接入点时，将服务进行迁移，这样将导致更高的延时，甚至是服务中断。由于用户的移动性，在服务请求次数高于25次时，延时抖动比较厉害。It is obvious that the delay was the highest when there was no migration scheme.

Fig. \ref{fig008} shows the response time of the edge node providing the emotion detection service with low resolution (the first algorithm introduced in Table \ref{tab.acc}). In general, the response time is proportional to the number of concurrent requests, which means that the more the number of concurrent requests is, the longer the average response time is.
We compare the time delay under different schemes, from Fig. \ref{fig008}, we can obtain that the delay of all the schemes increases with the increase of the service request times, our proposed RL-based scheme under ECC architecture has better performance with the lowest latency. This is because these services could be better migrated in advance to
%the nearest edge node to the user's request access point
the optimal location
based on a user-mobility prediction. However, the nearest migration scheme decides to migrate the service just when the user moves to the other access point, which would result in a longer delay and even to lead service disruption.
From Fig. \ref{fig008} we can also observe that due to user mobility, the delay jitter is more severe and the delay difference of these three schemes is bigger when the service request time is higher than 25. The ECC-based scheme is able to reduce the jitter from the results for it learns from the user mobility continuously. Also, it is obvious that the delay was the longest under no migration scheme.

%图1(提供不同service resolution的累计奖励对比图)，刚开始不同service resolution获得的收益差别不大，到稳定阶段，high resolution 获得reward 更大。
%
%%图2(reward学习过程中，利用移动性预测信息和不利用的对比图)，
%
%图2(时延分析)
%
%图3(emotion detection and video streaming services 对比图)

%Experimental results show that our proposal can reduce the migration cost up to xx compared to the xx scheme.

%-------------------------------
\section{Conclusions}
%-------------------------------
This paper presents an ECC network architecture and introduces the key issues. In addition, an ECC platform for dynamic service migration based on a mobile user's behavioral cognition was developed and experimentally tested. The experimental results  show that the proposed ECC architecture can simultaneously provide higher QoE compared with the general edge-computing architecture without data and resource cognitive engines that achieve the user behaviour prediction to better guide the service migration based on traffic data and the network resource environment. The results effectively demonstrated that edge cognitive computing realizes the cognitive information cycle for human-centered reasonable resource distribution and optimization.

\begin{acks}
This work is supported by  the National Natural Science Foundation of China
(Grant No. 61802138, 61802139, 61572220). This work has also been partially carried out under the framework of INTER-IoT, Research and Innovation action - Horizon 2020 European Project, Grant Agreement No. 687283, financed by the European Union.
Dr. Humar would like to acknowledge the financial support from the Slovenian Research Agency (research core funding No. P2-0246). Dr. Yixue Hao is the corresponding author.
\end{acks}

% Bibliography
\bibliographystyle{ACM-Reference-Format}

%\bibliography{sample-bibliography}

\begin{thebibliography}{1}

\bibitem{Sarros2018edge}
C. A. Sarros et al., ``Connecting the Edges: A Universal, Mobile-Centric, and Opportunistic Communications Architecture,'' \emph{IEEE Communications Magazine}, vol. 56, no. 2, pp. 136-143, Feb. 2018.

\bibitem{wei2016}
W. Shi, J. Cao, Q. Zhang, Y. Li, L. Xu. ``Edge Computing: Vision and Challenges,'' \emph{IEEE Internet of Things Journal}, vol.~3, no.~5, pp. 637--646, 2016.

\bibitem{eciot}
O. Salman, I. Elhajj, A. Kayssi, A. Chehab. ``Edge computing enabling the Internet of Things,'' \emph{Internet of Things IEEE}, pp. 603--608, 2016.

\bibitem{agent16}
M. Chen, Jun Yang , X. Zhu, X. Wang, M. Liu, J. Song, ``Smart Home 2.0: Innovative Smart Home System Powered by Botanical IoT and Emotion Detection'', \emph{Mobile Networks and Applications}, Vol. 22, pp. 1159-1169, 2017.

\bibitem{gian15}
G. Fortino, R. Gravina, W. Russo, C. Savaglio, ``Modeling and Simulating Internet-of-Things Systems:
A Hybrid Agent-Oriented Approach,'' \emph{Computing in Science and Engineering}, vol. 19, no. 5, pp. 68-76,
2017.

\bibitem{gian14}
M. Chen, Y. Miao, Y. Hao, K. Hwang, ``Narrow Band Internet of Things'', \emph{IEEE Access}, Vol. 5, pp. 20557-20577, 2017.

\bibitem{Edge18IIoT}
M. Chen, Y. Hao, ``Task Offloading for Mobile Edge Computing in Software Defined Ultra-dense Network'', \emph{IEEE Journal on Selected Areas in Communications}, Vol. 36, No. 3, pp. 587-597, Mar. 2018.

\bibitem{edge17rana}
I. Petri, O.F. Rana, J. Bignell, S. Nepal and N. Auluck. ``Incentivising Resource Sharing in Edge Computing Applications,'' pp. 204--215, 2017.

\bibitem{iot17singh}
Y. Qian, M. Chen, J. Chen, M. Hossain, A. Alamri, ``Secure Enforcement in Cognitive Internet of Vehicles'', \emph{IEEE IoT Journal}, Vol. 5, No. 2, pp. 1242-1250, 2018.

\bibitem{secu17rana}
M. Villari, M. Fazio, S. Dustdar, O. Rana, L. Chen, R. Ranjan. ``Software Defined Membrane: Policy-Driven Edge and Internet of Things Security,'' \emph{IEEE Cloud Computing}, vol.~4, no.~4, pp. 92--99, 2017.

\bibitem{zhou17}
L. Zhou, D. Wu, Z. Dong, and X. Li, ``When Collaboration Hugs Intelligence: Content Delivery over Ultra-Dense Networks,'' \emph{IEEE Communications Magazine}, vol. 55, no. 12, pp. 91-95, 2017.

\bibitem{zhou18}
L. Zhou, D. Wu, J. Chen, and Z. Dong, ``Greening the Smart Cities: Energy-Efficient Massive Content Delivery via D2D Communications,'' \emph{IEEE Transactions on Industrial Informatics}, vol. 14, no. 4, pp. 1626-1634, Apr. 2018.

\bibitem{soft18}
H. Habibzadeh, A. Boggio-Dandry, Z. Qin, T. Soyata, B. Kantarci and H. T. Mouftah, ``Soft Sensing in Smart Cities: Handling 3Vs Using Recommender Systems, Machine Intelligence, and Data Analytics,'' \emph{IEEE Communications Magazine}, vol. 56, no. 2, pp. 78-86, Feb. 2018.

\bibitem{cog18city}
M. Mohammadi and A. Al-Fuqaha, ``Enabling Cognitive Smart Cities Using Big Data and Machine Learning: Approaches and Challenges,'' \emph{IEEE Communications Magazine}, vol. 56, no. 2, pp. 94-101, Feb. 2018.

\bibitem{deep2018a}
A. Abeshu and N. Chilamkurti, ``Deep Learning: The Frontier for Distributed Attack Detection in Fog-to-Things Computing,'' \emph{IEEE Communications Magazine}, vol. 56, no. 2, pp. 169-175, Feb. 2018.

\bibitem{learnEdge}
M. Chen, V. Leung, ``From Cloud-based Communications to Cognition-based Communications: A Computing Perspective'', \emph{Computer Communications}, Vol. 128, pp. 74-79, 2018.

\bibitem{deep18veh}
Y. He, N. Zhao and H. Yin, ``Integrated Networking, Caching, and Computing for Connected Vehicles: A Deep Reinforcement Learning Approach,'' \emph{IEEE Transactions on Vehicular Technology}, vol. 67, no. 1, pp. 44-55, Jan. 2018.

\bibitem{sdn17deep}
M. Chen, Y. Hao, M. Qiu, J. Song, D. Wu, I. Humar, ``Mobility-aware Caching and Computation Offloading in 5G Ultradense Cellular Networks'', \emph{Sensors}, Vol. 16, No. 7, pp. 974-987, 2016.

\bibitem{alex17}
S. Zhang, S. Zhang, T. Huang, et al. ``Speech Emotion Recognition Using Deep Convolutional Neural Network and Discriminant Temporal Pyramid Matching,'' \emph{IEEE Transactions on Multimedia}, Vol. 20, No. 6, pp. 1576 - 1590, Oct. 2017.

\bibitem{vgg14}
M. Chen, X. Shi, Y. Zhang, D. Wu, M. Guizani, ``Deep Features Learning for Medical Image Analysis with Convolutional Autoencoder Neural Network'', \emph{IEEE Trans. Big Data}, DOI:10.1109/TBDATA.2017.2717439, 2017.

\bibitem{mig17}
A. Machen, S. Wang, K. Leung, et al. ``Live Service Migration in Mobile Edge Clouds,'' \emph{IEEE Wireless Communications}, Vol. 25, No. 1, pp. 140-147, 2017.

\bibitem{mig14}
V. Medina, J. Garcia, ``A survey of migration mechanisms of virtual machines,'' \emph{ACM Computing Surveys}, vol. 46, no. 3, pp. 30-62, 2014.

\bibitem{RL}
K. Hwang, M. Chen, ``Big Data Analytics for Cloud/IoT and Cognitive Computing,'' \emph{Wiley}, U.K., ISBN: 9781119247029, 2017.

\end{thebibliography}

\end{document}